# Enjeux et propositions sur les architectures RF pour l'homme connecté à la société numérique

Guillaume Villemaud*, Florin Hutu*, Tanguy Risset*, Jean-Marie Gorce*

*Université de Lyon, INRIA, INSA-Lyon, CITI-INRIA, 6, avenue des Arts, F-69621, Villeurbanne, {guillaume.villemaud}@insa-lyon.fr

**Résumé**
Cet article présente une vue générale des enjeux du développement croissant de liaisons sans fil visant à permettre une interconnexion de plus en plus constante et transparente entre l'homme et le monde numérique. Ces enjeux se situent au niveau des architectures de terminaux hautes performances pour les applications classiques de l'Internet sans fil, mais également dans le domaine des réseaux de capteurs et des objets connectés. Au-delà, les contraintes de ces diverses applications poussent à développer des architectures à forte composante numérique de type radio logicielle. Dans chacune de ces catégories, des exemples d'approches proposées par l'équipe Inria Socrate sont présentées.

**Abstract**
This article presents an overview of the challenges of increasing development of wireless links to enable a more consistent and transparent interconnection between people and the digital world. These issues are in the domain of high performance architectures for conventional applications of wireless Internet, but also in the field of sensor networks and connected objects. Beyond the constraints of the various applications push to develop architectures with high digital capabilities like software defined radio. In each of these categories, examples of approaches proposed by INRIA Socrate team are presented.

**Introduction**
Les évolutions des mœurs et des usages de notre société, en lien avec les progrès technologiques des communications sans fil, permettent aujourd'hui d'offrir au plus grand nombre un lien quasi permanent entre l'homme et son double numérique. Cette interconnexion constante nécessite de nouvelles approches et de nouveaux développements pour offrir une expérience la plus transparente possible, une qualité de service optimale en fonction du contexte, ainsi qu'un coût financier et énergétique le plus raisonnable possible. Nous présentons dans cet article quelques exemples de travaux développés au sein de l'équipe Inria Socrate basée à l'INSA de Lyon permettant d'essayer de répondre à ces enjeux. Dans la première section, les enjeux liés au développement de terminaux haute performance possédant de multiples degrés de liberté (dits multi-*) sont illustrés au travers de deux thèmes : les récepteurs multi-* et les relais multi-*. La section suivante traite de compromis dans le cadre des réseaux de capteurs et des réseaux domestiques : les techniques de Wake-Up radio ainsi que les approches de full-duplex. Enfin, la dernière section développes les approches basées sur la radio logicielle pour offrir plus de souplesse à ces systèmes, c'est-à-dire plus de flexibilité aux terminaux et la possibilité de reconfiguration entièrement logicielle.

## 1. Terminaux hautes performances : les architectures multi-*

Une des voies les plus évidentes d'évolution des communications sans fil et la convergence de nombreux standards au sein d'un seul et même terminal. Aujourd'hui, cette convergence s'exerce principalement au niveau des *smartphones* qui permettent une connectivité quasi-permanente par la combinaison de multiple standards : normes de téléphonie (2, 3 ou 4G), normes WLAN (WiFi a/b/g/n/ac), ou normes pour objets connectés (Bluetooth, ZigBee, UWB, NFC…). Mais ces systèmes utilisent pour cela une multiplication de puces radios, chacune dédiée à l'un de ces standards de communications. Cette approche n'étant optimale ni du point de vue coût, ni du point de vue consommation, ni même d'un point de vue encombrement, il apparaît de plus en plus pertinent de concevoir intrinsèquement ces terminaux comme étant multi-standard, et donc de chercher à mutualiser au mieux les composants nécessaires à ces différents standard. De plus, la plupart des évolutions des standards requérant des modes de communication multi-antenne ou également l'utilisation de bandes fréquentielles fragmentées, ces terminaux doivent tendre vers un nouveau paradigme de développement : celui des architectures dites multi-*. Nous allons donc présenter ici deux illustrations de cette approche : la conception de récepteurs multi-*, et au-delà le potentiel d'utilisation de tels terminaux comme points de relais dans un réseau large échelle : les relais multi-*.

## 1.1 Récepteurs multi-*

La mutualisation des composants dans les architectures de frontaux RF est un enjeu crucial pour réduire la consommation énergétique et donc également l'autonomie de ces équipements. Si intrinsèquement un terminal est dédié à travailler sur plusieurs standards de communications simultanément, cette approche apparait indispensable. Mais au-delà, l'intégration de traitements multi-antenne peut permettre, en plus du classique gain de performance sur le lien radio, de profiter des degrés de liberté supplémentaires offerts pour améliorer la réutilisation spectrale, ou encore permettre l'utilisation de composants aux contraintes relâchées (approche de *Dirty RF*). Dans ce cadre, plusieurs architectures de récepteurs multi-* ont été proposées. Tout d'abord, [1] présente le développement d'un récepteur multi-canal, multi-mode et multi-antenne permettant la numérisation simultanée de plusieurs canaux dans la bande à 2.45 GHz, offrant un potentiel de réjection spatiale et de réutilisation de canaux recouvrants. [2] présente l'étude d'un récepteur original à double translation orthogonale pour la réception de deux bandes de fréquences en simultané (voir Figure 1). Au-delà, on peut également citer la proposition d'architecture pour des terminaux de type LTE-Advanced qui offrent la capacité de travailler sur deux sous-bandes fractionnées avec une architecture nativement multi-antenne, avec une réduction de consommation évaluée à 33% par rapport à une architecture classique [3].

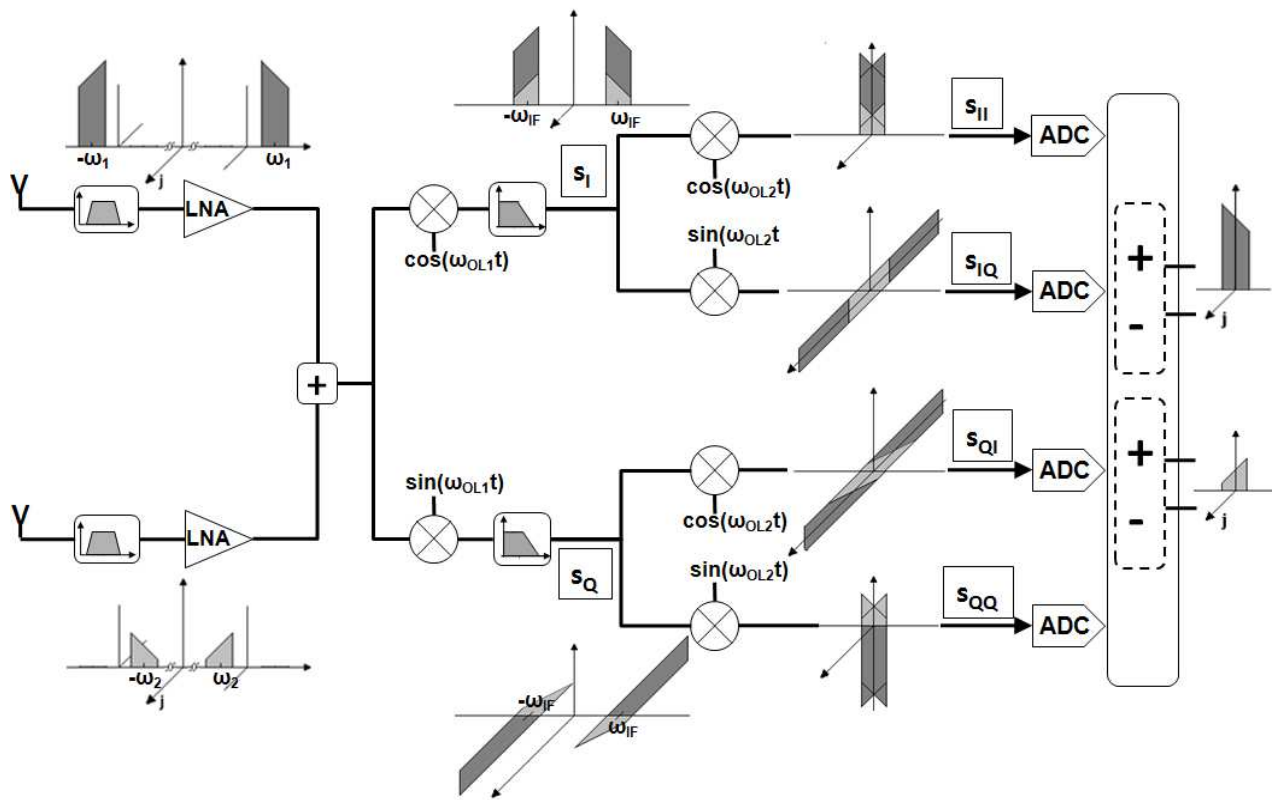

*Figure 1.      Principe de l'architecture double IQ pour la réception simultanée de deux bandes.*

## 1.2 Relais multi-*

Pour accroitre encore la disponibilité des liaisons radio pour l'utilisateur, quelque soit l'environnement et la proximité des points d'accès ou stations de base servant de passerelle avec le monde numérique, en plus des liens directs avec ces passerelles, il est intéressant de profiter du relayage potentiel de l'information par les objets communicants présents dans la cellule de communications, qu'ils soient fixes ou mobiles. Cette approche, déjà très éprouvée dans le domaine des réseaux de capteurs ou des réseaux ad hoc, peut naturellement offrir de nouvelles perspectives si les nœuds radios effectuant ce relayage sont intrinsèquement multi-*. En effet, le fait de passer le signal à relayer sur un autre standard de communication peut permettre de profiter de canaux de communication éventuellement plus favorables que ceux du réseau primaire, mais également de privilégier le standard le moins énergivore pour cette opération. Néanmoins, il a été montré dans [4] que ce choix de standard basse consommation n'apporte pas nécessairement de gain notable en fonction des scénarios considérés. Malgré cela, la considération d'architecture multi-antenne permet, pour des réseaux denses, d'offrir des potentiels nettement plus intéressants de ces relais multi-* [5].

## 2. Réseaux de capteurs et réseaux domestiques: des compromis taux d'utilisation - consommation à définir

Idéalement, l'homme connecté doit l'être partout et à tout moment. Cependant, maintenir une liaison radio en permanence n'est pas toujours nécessaire, quand aucun transfert de données n'est requis. Dans le cadre des réseaux domestiques par exemple, la passerelle n'a pas besoin de maintenir la connexion avec tous les équipements de l'habitation tout au long de la journée. C'est pourquoi les techniques de réveil des équipements à la demande (Wake-Up radio) se sont popularisées récemment. A l'inverse, dans certains scénarios de réseaux de capteurs à forte connectivité (réseaux sur le corps humains par exemple), des échanges constants d'information peuvent être requis, même si ces communications ne nécessitent pas toujours de forts débits. De nombreuses propositions de mécanismes de relais dans ces réseaux s'appuient sur la présupposition d'une interface de communication *Full-Duplex*, à savoir capable d'émettre et recevoir au même instant à la même fréquence.

### 2.1 Réseaux domestiques : approches de Wake-Up radio

Différentes approches Wake-Up existent, basées sur des mécanismes d'accès au médium (contrôle MAC des temps d'éveil), ou au niveau de la couche physique (PHY). Les mécanismes de la couche PHY peuvent s'appuyer sur des modules radios additionnels entièrement passifs (type rectenna par exemple), mais ces modules ont le défaut important de provoquer de nombreux faux réveils et d'avoir une portée généralement limitée. Mais des modules actifs faible consommation existent, permettant de créer des signaux de réveil avec un identifiant ciblant exactement l'objet communicant à éveiller. Le compromis à trouver dans toutes les approches proposées réside entre la complexité de l'architecture additionnelle mise en œuvre et sa consommation. Plus le réseau a un taux d'utilisation (*duty cycle*) faible, plus l'apport de ces mécanismes sera intéressant. Au-delà, plus le réseau considéré est dense, plus le recours à des modules permettant l'identification sera à privilégier. Par exemple dans [6], une stratégie d'identification de modules WiFi à partir de l'émission de signaux OFDM à empreinte fréquentielle particulière permet le recours à des modules Wake-Up très faible consommation avec une identification efficace (voir Figure 2).

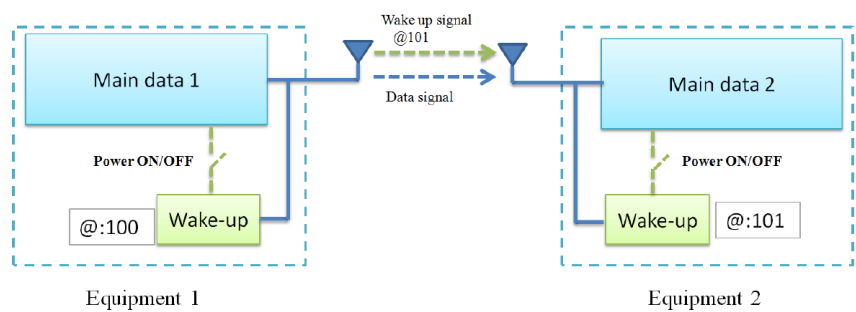

*Figure 2.    Principe du module de Wake-Up avec identification.*

### 2.2 Réseaux de capteurs à fort duty-cycle : l'approche Full-Duplex

A l'inverse des approches Wake-Up, les systèmes Full-Duplex visent à offrir une connectivité optimale à tout moment. En effet tous les systèmes communicants actuels partages le medium radio en temps ou en fréquence. Essayer de recevoir un signal alors même que l'on est en train d'émettre au même instant à la même fréquence pose le problème de la très forte auto-interférence que l'on crée, dans laquelle le signal d'intérêt est complètement noyé. Pour permettre de combiner l'émission et la réception, les terminaux full-duplex tirent parti de la connaissance du signal qu'ils envoient et qui est source de l'auto-interférence. Une estimation fine du canal radio de l'auto-interférence permet alors d'appliquer des algorithmes de suppression de cette interférence. Cette suppression peut être faite au niveau analogique, numérique ou encore aux deux niveaux. Une approche numérique pure est néanmoins assez contrainte, car dans ce cas la numérisation du signal va se faire à l'échelle du signal interférent, et dès lors l'information résiduelle du signal d'intérêt souffre d'une très faible résolution. Dans [7] par exemple, une suppression au niveau analogique est appliquée à partir d'une estimation du canal faite en numérique individuellement pour chaque sous-porteuse d'un signal OFDM 802.11 (Figure 3).

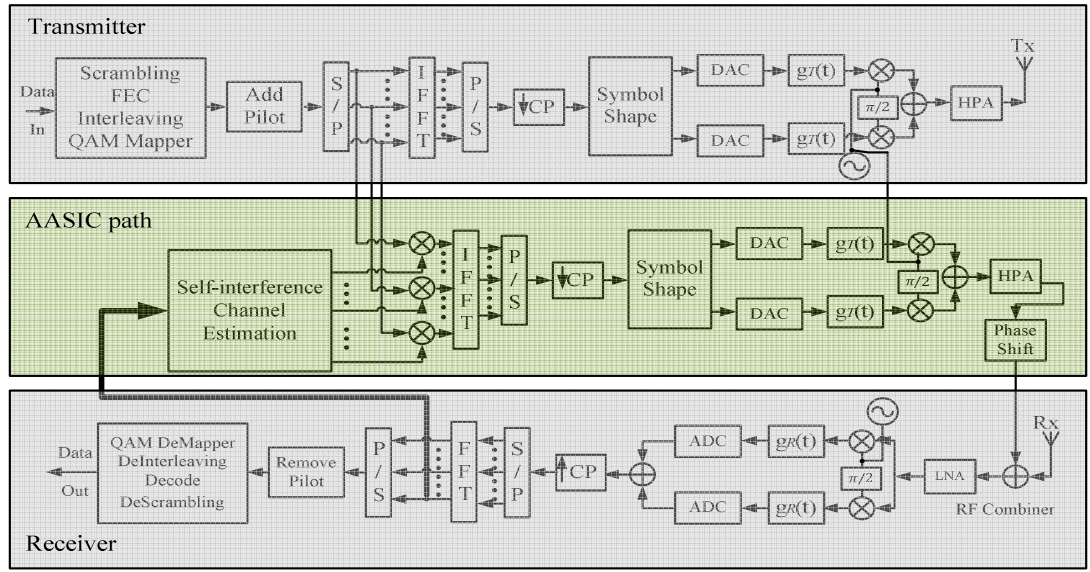

*Figure 3.       Schéma de principe du système Full-Duplex OFDM.*

## 3. Flexibilité et reconfigurabilité: le potentiel des approches par radio logicielle

Qu'un système offre des degrés de liberté multiples, qu'il permette des communications simultanées, cela permet déjà de répondre à de nombreux challenges liés à l'homme connecté. Mais cette flexibilité n'a vraiment de potentiel que si elle est associée à des capacités de traitement numérique suffisantes. Et pour aller au-delà, et surtout laisser la porte ouverte à des possibilités d'évolution, c'est-à-dire de reprogrammation des interfaces logicielles. Le concept de radio logicielle définie, ou *Software Defined Radio* (SDR) a été édicté en ce sens, et a donné lieu à de très nombreux développement dans les années passées, que ce soit pour des systèmes opérationnels mais aussi et surtout pour de puissants systèmes de tests et développements. Nous citons ici un exemple d'étude du recours à ces systèmes SDR pour les réseaux de collectes en milieu urbain, puis l'exemple d'une plateforme d'envergure basée sur des modules radio logicielle permettant notamment l'étude des approches de radio cognitive.

### 3.1 Points de collecte dans les réseaux urbains : passerelles SDR

Les applications de relevé d'information en milieu urbain augmentent au fil des ans, à mesure justement que l'homme se retrouve de plus en plus connecté au monde numérique. Mais cette augmentation s'accompagne également du développement de nombreux standards dédiés à ces applications, technologies normalisées ou propriétaires. Dès lors, les fournisseurs de services associés ou plus largement les opérateurs ont tout intérêt à développer des puits de collecte les plus flexibles et évolutifs possibles. La bande fréquentielle majoritairement dédiée à ces applications n'étant de plus pas très large (8 MHz pour la bande ISM à 868 MHz), il devient envisageable de concevoir des passerelles SDR numérisant l'ensemble de la bande pour recevoir simultanément différents signaux sur différents canaux, également potentiellement avec différents standards de communication. Cependant, pour des réseaux à très large échelle pouvant regrouper plusieurs milliers de nœuds par passerelle, les dynamiques de signaux qui peuvent cohabiter deviennent rapidement prohibitives. [8] montre que dans un scénario de type SmartSantander [9], la dynamique conduit à une résolution nécessaire pour l'échantillonnage simultané de plusieurs signaux d'au moins 21 bits, ce qui n'est à l'heure actuel pas envisageable.

### 3.2 Plateforme d'expérimentation large échelle de radio cognitive: CorteXlab

Enfin, une grande nouveauté apportée par les technologies SDR est la possibilité d'accélérer grandement les cycles de conception et d'expérimentation. En effet, une architecture SDR associée à un front-end RF suffisamment flexible permet de mettre en place très rapidement des tests pour mettre à l'épreuve les futurs standards de communication, les nouvelles stratégies d'allocation de ressources, ou les nouvelles approches de coopération ou de relai par exemple. De plus en plus de solutions commerciales existent pour émuler la plupart des standards de communication existants (voir [10]). Particulièrement, les approches de radio cognitive, qui visent à occuper plus dynamiquement le spectre radio en cherchant continuellement les fréquences non occupées pour émettre, prennent tout leur sens avec ces architectures SDR. Mais dans un contexte fortement hétérogène de réseaux denses entremêlés, les approches théoriques peinent à prédire fidèlement la capacité de ce type de liens radios. La plateforme CorteXlab [11], composée de nombreux boîtiers SDR cohabitant dans un large environnement confiné et tous programmables à distance, permet de lancer de nombreux cycles de tests de ces approches novatrices. Cette plateforme prévoyant de plus l'ajout de nœuds mobiles robotisés, la plupart des scénarios liés au futur de l'homme connecté pourront y être évalués et analysés.

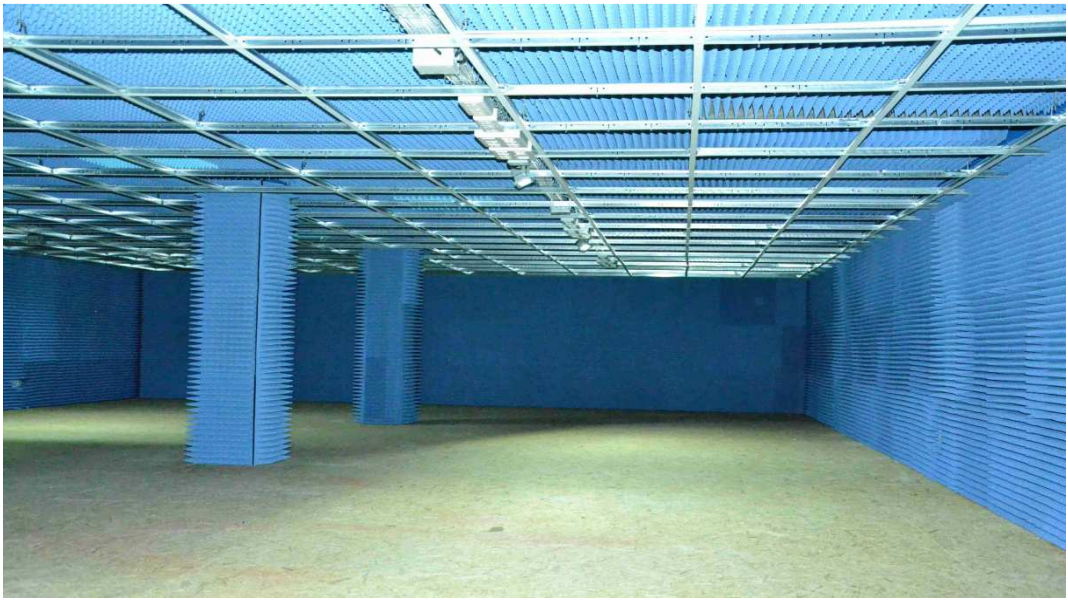

*Figure 4.     La salle d'expérimentation de la plateforme CorteXlab à l'INSA de Lyon.*

### Références